# Astro2020 Science White Paper

# A Direct Measure of Cosmic Acceleration

**Thematic Areas:** ☐ Planetary Systems  ☐ Star and Planet Formation
☐ Formation and Evolution of Compact Objects  ☒ Cosmology and Fundamental Physics
☐ Stars and Stellar Evolution  ☐ Resolved Stellar Populations and their Environments
☐ Galaxy Evolution  ☐ Multi-Messenger Astronomy and Astrophysics

**Principal Authors:**
Name: Stephen Eikenberry & Anthony Gonzalez
Institution: University of Florida

**Co-authors:** (names and institutions)
Jeremy Darling (University of Colorado), Zachary Slepian (University of Florida), Guido Mueller (University of Florida), John Conklin (University of Florida), Paul Fulda (University of Florida), Sarik Jeram (University of Florida), Chenxing Dong (University of Florida), Amanda Townsend (University of Florida)

**Abstract** (optional):
Nearly a century after the discovery that we live in an expanding Universe, and two decades after the discovery of accelerating cosmic expansion, there remains no direct detection of this acceleration via redshift drift — a change in the cosmological expansion velocity versus time. Because cosmological redshift drift directly determines the Hubble parameter H(z), it is arguably the cleanest possible measurement of the expansion history, and has the potential to constrain dark energy models (e.g. Kim et al. 2015). The challenge is that the signal is small — the best observational constraint presently has an uncertainty several orders of magnitude larger than the expected signal (Darling 2012). Nonetheless, direct detection of redshift drift is becoming feasible, with upcoming facilities such as the ESO-ELT and SKA projecting possible detection within two to three decades. This timescale is uncomfortably long given the potential of this cosmological test. **With dedicated experiments it should be possible to rapidly accelerate progress and detect redshift drift with only a five-year observational baseline.** Such a facility would also be ideal for precision radial velocity measurements of exoplanets, which could be obtained as a byproduct of the ongoing calibration measurements for the experiment.

**Motivation:** During the past few decades, precision astrophysical observations have led to the realization that we live in a Universe dominated by dark energy and dark matter. This revolutionary new view of the "Dark Cosmos" drives both our understanding of the Universe and our roadmap for investigating its contents and fundamental laws. However, while dark energy can be accommodated as a cosmological constant in General Relativity, the Standard Model of particle physics predicts an amplitude for the cosmological constant that differs by a factor of $10^{120}$ – perhaps the greatest conflict in all of modern physics. This failure has motivated numerous theoretical models to explain dark energy (see e.g Copeland, Sami, Tsuijkawa (2006), for a review) and has spurred a series of observational programs dedicated to testing this picture (e.g. Albrecht et al. 2009, joint report of the Dark Energy Task Force). Meanwhile there are numerous large experiments and space missions underway to better constrain the dark energy equation of state, parametrized by $w$ = pressure/energy density, and its evolution. Examples include the Dark Energy Survey (DES), the Dark Energy Spectroscopic Instrument (DESI), Euclid, and the Wide Field InfraRed Space Telescope (WFIRST), and most major upcoming facilities include dark energy as one of their key science cases (e.g. the Large Synoptic Survey Telescope – LSST).

It is important to realize that the existing techniques for testing dark energy each have implicit assumptions (i.e. the cosmological distance ladder and/or calibrations of various distance indicators), and consequently none provide a model-independent determination of the expansion history of the Universe. Furthermore, at a fundamental level it is the expansion history that encodes the net influences of dark energy and dark matter over cosmic time. One can in principle directly determine H(z) by observing the change in velocity of a receding object over time. This is nothing more than using the basic definition of acceleration: $a = dv/dt \cong c\, dz/dt$. As first demonstrated by McVittie (1962) and Sandage (1962), the relation between this acceleration and H($z$) is simply H($z$) = (1+$z$) H$_0$ - d$z$/d$t$ (Figure 1). Direct measurement of d$z$/d$t$ can distinguish between a "coasting" (empty) Universe and a Universe containing matter, radiation, dark energy, etc. (Figure 2). Such a direct, simple approach is crucial for verifying our current paradigm and providing unique, model-free constraints that complement existing approaches. The need for such a comparison is more true now than ever given the recent tension between supernovae and CMB H$_0$ measurements (e.g. Riess et al. 2016, Bernal et al. 2016). If real, this tension might perhaps argue for new physics (e.g. Mörtsell & Dhawan 2018, Kreisch et al. 2019); it cannot be explained by merely local sample variance (Wu & Huterer 2017). However, another possible cause of the tension is observational systematics (Efstathiou 2014; Addison et al. 2016). In either case, a new, independent probe of basic cosmological parameters will be extremely valuable.

**Observable Physical Systems and the State of the Field**

Detection of redshift drift requires redshift measurements of spectral features that, at least in an ensemble average, are stable over the period of the experiment and have relatively low intrinsic linewidths. The two most promising classes of objects for such observations are 21 cm emission and the Ly$\alpha$ forest. In the former case, the idea is to observe absorption lines from HI associated with galaxies at different epochs and either individually or statistically map the redshift drift. This approach has the advantage that one can in principle detect lines at any redshift.

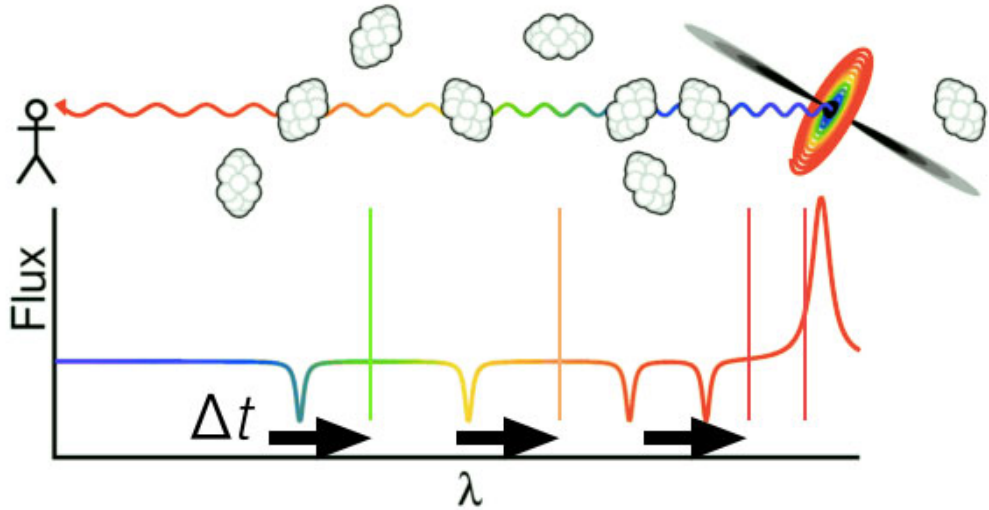

*Figure 1* – Conceptual cartoon of how redshift drift works. By monitoring absorption features in quasar spectra, caused by clouds of neutral gas at different z along the line of sight, we can detect shifts in z over time at different redshifts. This is a direct measurement of dz/dt. (Adapted from E. Wright web site http://www.astro.ucla.edu/~wright/Lyman-alpha-forest.html)

Use of the Lyα forest to measure redshift drift was proposed by Loeb (1998). The idea in this case is to use quasars as backlights and measure the drift of the Lyα forest lines relative to one another. While this approach is restricted to redshifts at which Lyα is observable, a single observation will contain many lines that can be used to infer d$z$/d$t$ as a function of redshift over the observable window. The fundamental challenge for measuring cosmological redshift drift is that the signal is small. Detection requires measuring redshifts with an accuracy of a few parts in $10^{10}$, or equivalently being able to detect velocity changes at the level of ~1 *cm/s/yr*, with an instrument that is stable at this precision over the timescale of the measurement (see Figure 2). The most stringent current upper limit on cosmological redshift drift is from Darling (2012). Using 21 cm absorption line observations at *z*=0-1, this study found d$z$/d$t$ = (−2.3 ± 0.8) × $10^{-8}$ $yr^{-1}$ or Δ$v$/Δ$t$ = −5.5 ± 2.2 *m s$^{-1}$ yr$^{-1}$* . These uncertainties are still more than two orders of magnitude larger than the expected signal.

Much like LIGO, however, achieving the requisite precision is a matter of effort and engineering, and the requisite instrumental precision is within reach. Liske et al. (2008) make the case for measurement of redshift drift with upcoming 30m-class telescopes. This study found that a detection via the Lyα forest should be possible given observations separated by a 20-30 year baseline with a sufficiently stable spectrograph. This motivated a design study for the COsmic Dynamics and EXo-earth experiment (CODEX) for the European Extremely Large Telescope (E-ELT), a precursor to the now-approved second-generation instrument HIRES. Redshift drift is one of the key science cases for this instrument, which has completed its Phase A study and forecast to be on sky by the late 2020s. Given this timescale, a detection of redshift drift by

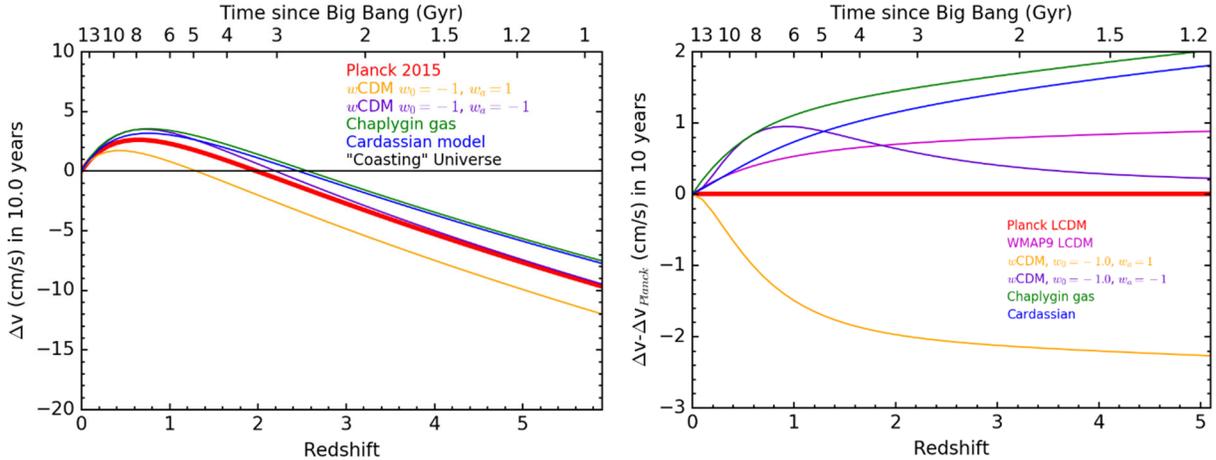

***Figure 2** – Expected redshift drift versus redshift over a 10-year measurement baseline. (Left) Expected total drift over 10 years. Note that a "coasting Universe" can be easily distinguished from models containing dark matter/energy, with a precision of ~cm/s. (Right) The same models, but now using a Planck cosmology as the reference for a standard LCDM cosmology (Planck Collaboration 2015), models with evolving w. We also plot two alternative cosmologies for illustration, a Chaplygin gas and Cardassian models. For these we use the best fit constraints derived from current data sets presented in Li, Yang, & Wu (2018), and Wang & Meng (2016), respectively. On a 10-year baseline, ~1 cm/s precision (as a function of redshift) will allow redshift drift to distinguish clearly between different cosmological models.*

~2050 is plausible. Figure 3 below illustrates the expected precision of constraints on the dark energy equation of state, *w*, the Hubble parameter, and the matter density, as well as the complementarity of these constraints to upcoming NASA/ESA dark energy missions.

In the radio regime, the upcoming general-use facility that is most promising for measuring redshift drift is the Square Kilometer Array (SKA). Klöckner et al. (2015) find that measuring cosmic acceleration is not possible with SKA Phase 1, but argue that with modifications to the baseline design SKA Phase 2 would be capable of observing redshift drift at low redshift with a ~12 years baseline. With the current SKA timeline projecting completion of Phase 2 construction by 2030 (https://unitedkingdom.skatelescope.org/ska-project/ska-timeline/) , this approach would yield a low-redshift detection no earlier than the early 2040s.

**The Advantage of Dedicated Experiments**

While these observatory-class facilities have the potential to detect cosmic acceleration by 2050, the US community does not have access to either facility and the timeline of ~30 years is untenable given currently pressing questions on the expansion history of the Universe and the role of dark energy. For these reasons, we advocate both for prioritizing this science within the US astronomical research portfolio and for consideration of dedicated cosmic dynamics experiments, which can be more cost-effective and deliver results on much shorter timescales than are achievable with these large facilities. The authors of this white paper include researchers active in exploring potential constraints with both optical and radio observations.

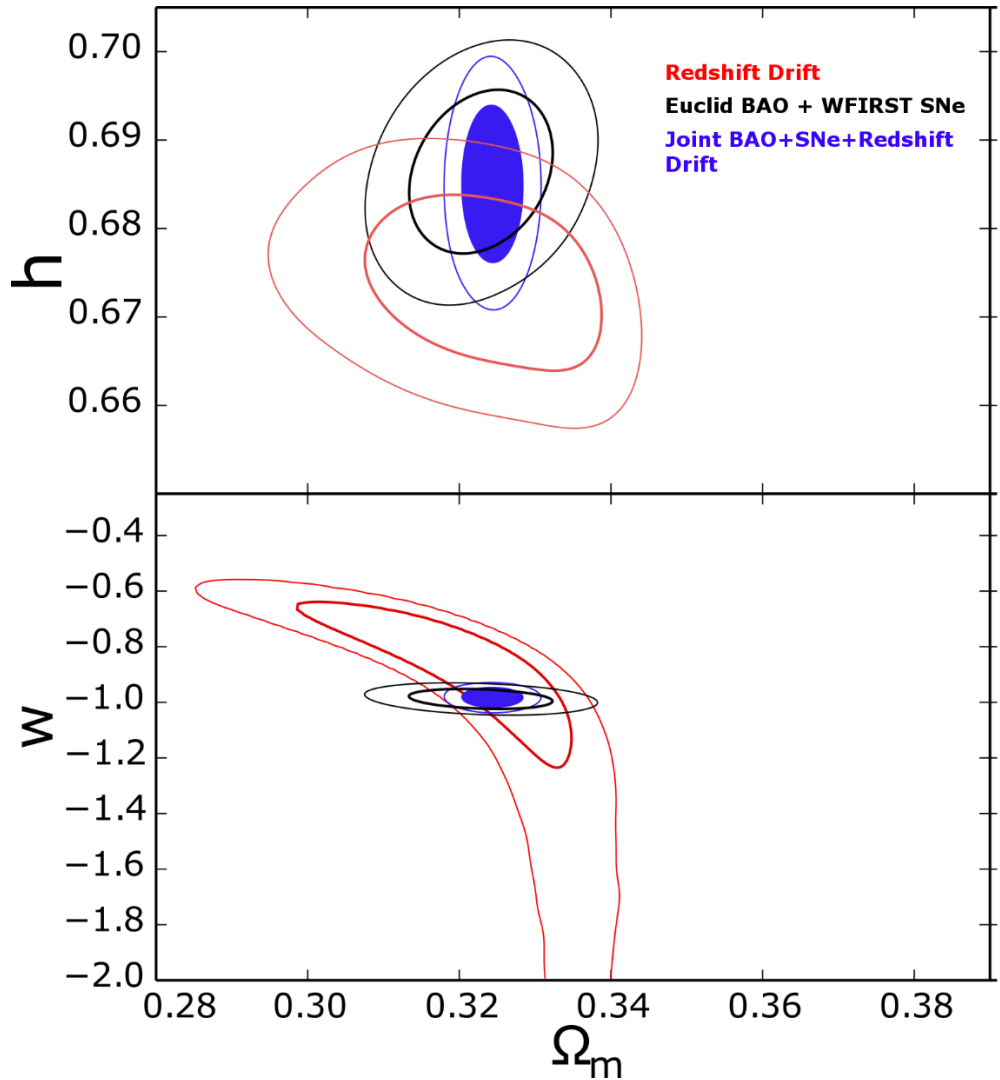

*Figure 3 - Future constraints on the Hubble parameter and the dark energy equation of state (w). (Based on Figure 5 of Lazkoz, Leanizbarrutia, & Salzano (2017)). The red contours show the expected redshift drift constraints from 24 years of observation using the approximate formula developed by Liske et al. (2008) for a CODEX-type experiment on the E-ELT. The black contours show the constraints expected from combining Euclid measurements of Baryon Acoustic Oscillations with WFIRST measurements of supernovae. The blue contours indicate joint constraints from combining the latter with redshift drift, emphasizing that the resulting uncertainties are significantly reduced by including redshift drift. **Critically, a dedicated experiment can achieve comparable precision for redshift drift in less than a decade.***

In the short term, we specifically advocate for dedicated optical experiments, with observing facilities devoted to making redshift drift measurements (as opposed to traditional facilities, like the E-ELT, which will only be able to commit a small fraction of their observing time to this). We have found that these offer the greatest potential to qualitatively advance the field given recent advances in actively stabilized high-resolution spectrographs and aperture synthesis for arrays of optical telescopes. Our team has calculated that for one such concept, a detection of dz/dt could be achieved within 5 years of operation (after a ~3-year build phase) using a dedicated facility that costs on the order of $100 million – quite cost-effective compared to the planned next-

generation dark energy missions and ground-based experiments (e.g. Euclid, WFIRST, CMB-S4). Such a facility would also be ideal for precision radial velocity measurements of exoplanets, which would be obtained as a byproduct of the ongoing calibration measurements for the experiment, enabling the community to address key science in two of the most active areas of astrophysics.

Generally speaking, any such redshift drift experiment shares three common needs. The first of these is large collecting area to provide the required photon flux for precision measurements of faint signals. Note that this does not necessarily imply high-cost diffraction-limited facilities such as the E-ELT — lower-performance (and lower-cost) "light buckets" composed of simplified telescopes, or arrays of linked smaller telescopes can meet the requirements for redshift drift measurement. Furthermore, the discovery of bright high-redshift quasars in current wide-field surveys can reduce the required collecting area to achieve a given precision in dz/dt. The second general requirement is a high-precision radial velocity spectrograph, capable of measurement precisions of ~ 1 cm/s. This has been the subject of significant technological development in recent years, largely driven by exoplanet science, and several approaches are currently close to achieving this performance level (e.g. Wilken et al., 2012). The third requirement is to stably calibrate the measurements to this level on timescales of decades. This requirement deviates from the current drivers for exoplanet science (where stability over weeks, months, or years is sufficient). One possibility would be to use an ensemble of stellar precision radial velocities as a long-term reference frame. While even quiet individual stars jitter at the level of ~10 cm/s, averages of tens or hundreds of such stellar velocities would be stable to <1 cm/s on long timescales. Thus, we conclude that the key technical requirements for a redshift drift measurement can be met in the immediate future.

**In summary, redshift drift experiments provide a unique complementarity to planned dark energy experiments and hold the promise of yielding the first direct measurement of the change in the expansion rate of the Universe. Current planned multi-use facilities capable of achieving a detection do not involve the US community and have timescales of 30-40 years until detection. We advocate for investment in dedicated experiments, which have the potential for a detection within 10 years, including construction, and can yield competitive dark energy constraints shortly thereafter. Such a facility also holds the promise of enabling ultra-high precision, longer term monitoring of exoplanet radial velocities and enabling other kinematic astrophysical measurements.**